\documentclass{aa}
\usepackage{graphicx}
\begin{document}
\title{A novel approach for extracting time-delays from lightcurves
of lensed quasar images}

\author{Ingunn Burud \inst{1,2} 
\and Pierre Magain \inst{1}
\and Sandrine Sohy  \inst{1}
\and Jens Hjorth  \inst{3}}

   \offprints{I. Burud}
\institute{Institut d'Astrophysique et de G\' eophysique, Universit\' e de
  Li\`ege, Avenue de Cointe 5, B--4000 Li\`ege, Belgium\\
\email{burud@astro.ulg.ac.be, magain@astro.ulg.ac.be, sohy@astro.ulg.ac.be}
\and
Space Telescope Science Institute, 3700 San Martin Drive, 
Baltimore, MD 21218, USA
\and
Astronomical Observatory, University of Copenhagen, 
Juliane Maries Vej 30, DK--2100~Copenhagen {\O}, Denmark \\
\email{jens@astro.ku.dk}
}

   \date{}

\abstract{ We present a new  method to estimate time delays from light
curves of lensed quasars. The method is based on $\chi^2$ minimization
between the data and a numerical model light curve. A linear variation
can be  included in order  to correct for slow  long-term microlensing
effects  in one  of the  lensed images.  An iterative  version  of the
method  can  be   applied  in  order  to  correct   for  higher  order
microlensing effects.  The method  is tested on simulated light curves. 
When  higher order microlensing effects are  present the time
delay is best constrained with the iterative method.  Analysis of 
a published data set for the lensed double Q~0957+561 yields results in
agreement  with other  published  estimates.  \keywords{Methods:  data
analysis -- Gravitational lensing }}

\maketitle
%

\section{Introduction}

The  time delay  between light  rays from  gravitationally  lensed quasar
images is a measurable parameter directly related to the gravitational
potential     and     to      the     Hubble     constant     $H_{0}$
(Refsdal~\cite{Refsdal}).  Accurate time delays obtained from multiply
lensed QSOs can  hence be used i) to  determine $H_{0}$ providing that
the  lens mass distribution  is known,  or ii)  to constrain  the mass
distribution in  a given lens,  once $H_{0}$ has been  determined from
other  methods  (or  other  lensed  QSO  systems).   Much  effort  has
therefore been devoted to observations  of lensed QSOs during the last
years, and in particular to the photometric monitoring of lensed QSO images.

Measuring the time  delay between the images of lensed quasars  is, for several reasons,
not a trivial  task. First it requires regular  monitoring of a target
over  a  long  period  (substantially  longer than  the  time  delay).
Second, the sampling is crucial and has
to be  determined on basis of  the intrinsic variations  in the quasar
and the estimated  time delay. Third, most objects  are not observable
during  the whole  year,  i.e., they  will  be below  the horizon  for
certain periods.  Fourth, there are nights with bad weather conditions
when no data are obtained.   Finally, variations in lensed quasars may
not only be  due to intrinsic fluctuations of the  quasar, but also to
microlensing  by compact  objects along  the  line of  sight.  Such  a
microlensing signal, depending on  its timescale and amplitude, can be
used  to constrain the  size of  the continuum  and the  line emitting
regions in  the quasar, and the  distribution of compact  matter in the
lens galaxy  (Paczy{\'n}ski ~\cite{Pacz}, Kayser  et al. ~\cite{Kayser}).   
However, as
long as these external variations are not clearly distinguished from
the intrinsic variations, microlensing remains a nuisance for time
delay measurements. For the above   reasons, 
 advanced statistical methods  have to be used  to measure time delays
from quasar light curves.

We have developed a method based  on $\chi^2$ minimization  of the data
and a numerical model light curve.  The aim was to develop a  method based on 
simple principles, and to be able to properly detect and correct for 
possible microlensing  effects.
It has already  been successfully
applied to several time delay measurements (Burud et al.~\cite{Burud};
Hjorth  et al.~\cite{Hjorth}).   In this
paper we will present the principles of the algorithm (Sect.~\ref{method})
apply it to various simulated light curves (Sect.~\ref{examples}) and
to a public dataset of QSO~0957+561 from Serra-Ricart  et al.  (\cite{Serra-Ricart}) (Sect.~\ref{sect:b0957}).

\section{Method}
\label{method}

Let us assume that we have light curves for two images, A and  B, of a
lensed quasar.  There are N data points in each of the time dependent
light curves  $a(t)$ and  $b(t)$ with measurement  errors $\sigma_{a}$
and $\sigma_{b}$.  The  two curves are identical except for a
shift  in time, $\Delta t $, and  in magnitude,
$\Delta m$. It is therefore possible  to model the two curves with one
model curve  $g(t)$, and the  parameters representing the  time delay
and the  magnitude shift.   In some  cases the light  from one  
or both of the
images may be microlensed by individual stars in
the  lensing  galaxy.  Except  for  the  cases  of high
amplification  events,  which  are  of  short  duration,  microlensing
effects are  often slow variations  that can be  modeled to first
order as a linear variation with slope $\alpha$.

As model  curve we  choose an  arbitrary curve with  a fixed  number of
equally spaced sampling points, $M$.  This model may be
$\chi^{2}$ minimized   to  the   two  observed  light   curves.   The
minimization is done only for the  N observed data points. In this way
only  the model  curve  is interpolated  and  not the  data.  We  thus
minimize the following function:

\begin{eqnarray}
{\cal F} & = & 
\sum_{i=1}^{N} \left[\frac{a(t_{i})-g(t_{i})}{\sigma_{a_{i}}}\right]^2
 \nonumber\\
& & + \sum_{i=1}^{N} \left[\frac{(b(t_{i}-\Delta t)-(\Delta m+\alpha t_{i}))-g(t_{i})
}{\sigma_{b_{i}}}\right]^2
\label{equation:min1}
\end{eqnarray}

The determination of  the time delay from sampled  light curves always
relies  on  the  implicit  assumption that  intrinsic  variations  are
continuous in  time and slow enough  to be measured,  given the adopted
frequency of the observations.   Consequently with this assumption, if
the  typical sampling  interval is  $\tau_1$, we  can thus  smooth the
model curve $g(t)$ on the  same time scale $\tau_1$ (e.g., typically 7
days when observations  are obtained once per week).   This is done by
introducing  a  smoothing  term  which  minimizes the  square  of  the
difference  between the original  model and  the same  model convolved
with a Gaussian $r$ whose Full Width at Half Maximum (FWHM) is $\tau_1$.
This smoothing is performed for all  the points $M$ on the model curve
$g(t)$ so that  the dates without data points are  smoothed out on the
model curve and gaps in  the observed light curves will not contribute
significantly to  the result.  The  smoothing term is multiplied  by a
Lagrange parameter  $\lambda$.  This parameter  can be chosen  so that
the  model curve  matches the  data correctly  in a  statistical sense
(i.e., we  get the correct $\chi^2$).  In  practice however, excessive
smoothing  will prevent the  minimization from  converging.  Therefore
the  smoothest  possible solution  which  allows  the minimization  to
converge is usually adopted.

In a $\chi^2$ minimization it is assumed that all
the data points are independent.  However, a light
curve often consists in groups of nearby points
and more isolated points.  In regions where the
time sampling is much shorter than the intrinsic
quasar variations, the different measurements do
not really bring new information on the shape of
the light curve.  Rather, they increase the
precision with which the quasar magnitude is
known at that moment.  They might even be affected
by common systematic errors, i.e. coming from a
microlensing variation with a time scale comparable
to the interval covered by these neighbouring points.

In such cases of strongly uneven time coverage, we
find that the best results are often obtained when a
weight $W_i$ is given to each data point in the
following manner. 
For  a point  at a  time $t_{i}$  we calculate  the  relative distance
$t_{i}-t_{j}$  to all  the  other  points in  the  curve.  A  Gaussian
function with a FWHM$=2\sqrt{2\ln (2)} \tau_{2}$ chosen by the user is
centered at  $t_{i}$. The inverse of  the sum of the  ordinates of the
Gaussian for each  of the points $t_{j}$ will give  the weight for the
point $t_{i}$.  The expression of $W_{i}$ can be written:
\begin{equation}
W_i = \frac{1}{\sum_{j=1}^{N} exp{(-\frac{t_{i}-t_{j}}{\tau_{2}})^2}}
\end{equation}
The weight is
normalized so that the maximum allowed weight for a data point is $1$, and
this will  occur only when one point is  present within the  time interval
defined by  $\tau_{2}$. The ideal
choice of  $\tau_{2}$ is the  approximate time scale of  variations in
the  quasar. For some systems  this time  scale will  be  long, e.g.,
several  hundred days, for  other systems  it can  be shorter  e.g., 50
days.  We can now write the final function to be minimized:

\begin{eqnarray}
{\cal F} & = &
\sum_{i=1}^{N} W_{i} \left[\frac{a(t_{i})-g(t_{i})}{\sigma_{a_{i}}}\right]^2 \nonumber\\
& & +\sum_{i=1}^{N} W_{i} \left[ \frac{(b(t_{i}-\Delta t)-
(\Delta m+\alpha t))-g(t_{i})}{\sigma_{b_{i}}}\right]^2  \nonumber\\
&  & +\lambda \sum_{i=1}^{M}\left[g(t_{i})-\left[r*g\right](t_{i})\right]^2
\label{equation:min2}
\end{eqnarray}
where $r*g$ signifies a convolution between $r$ and $g$.

\subsection{Microlensing effects and iterative version of the algorithm}

Slow microlensing variations in one of  the light curves are modeled as
a linear term  with the parameter $\alpha$.  If  higher order 
variations  are present we  can use  the method  in an  iterative way.
This  is done by  splitting the  light curves  into several  parts and
analyzing them separately.  The $\chi^2$ method
is applied to determine the amount of linear external variations 
($\alpha$) for each separate part and for a range of input time delay
values.  First we run the programme with the first  time
delay $dt_{1}$ in the chosen range and we estimate  $\alpha$ in each 
of  the separate
parts.  We then correct each  part with the corresponding $\alpha$ and
add  the   parts  back  together  in   order  to  obtain   a  set  of
microlensing-corrected light  curves for  the given input time delay $dt_{1}$.
We can  now run  the $\chi^2$ method  to measure the time  delay 
 for  these ``corrected''  curves.  This
procedure  is  repeated  for  all  the  input time  delay  values  $dt_{2}$,
$dt_{3}$, ...  $dt_{n}$ in the range  to be studied.   We finally
obtain a time delay measurement for each of the $n$ input values,
and all these  measurements will
generally converge towards the  real  time delay value.

\section{Time delay measurement on simulated light curves}
\label{examples}

\subsection{Data simulations}

\begin{figure}
\includegraphics[width=7cm]{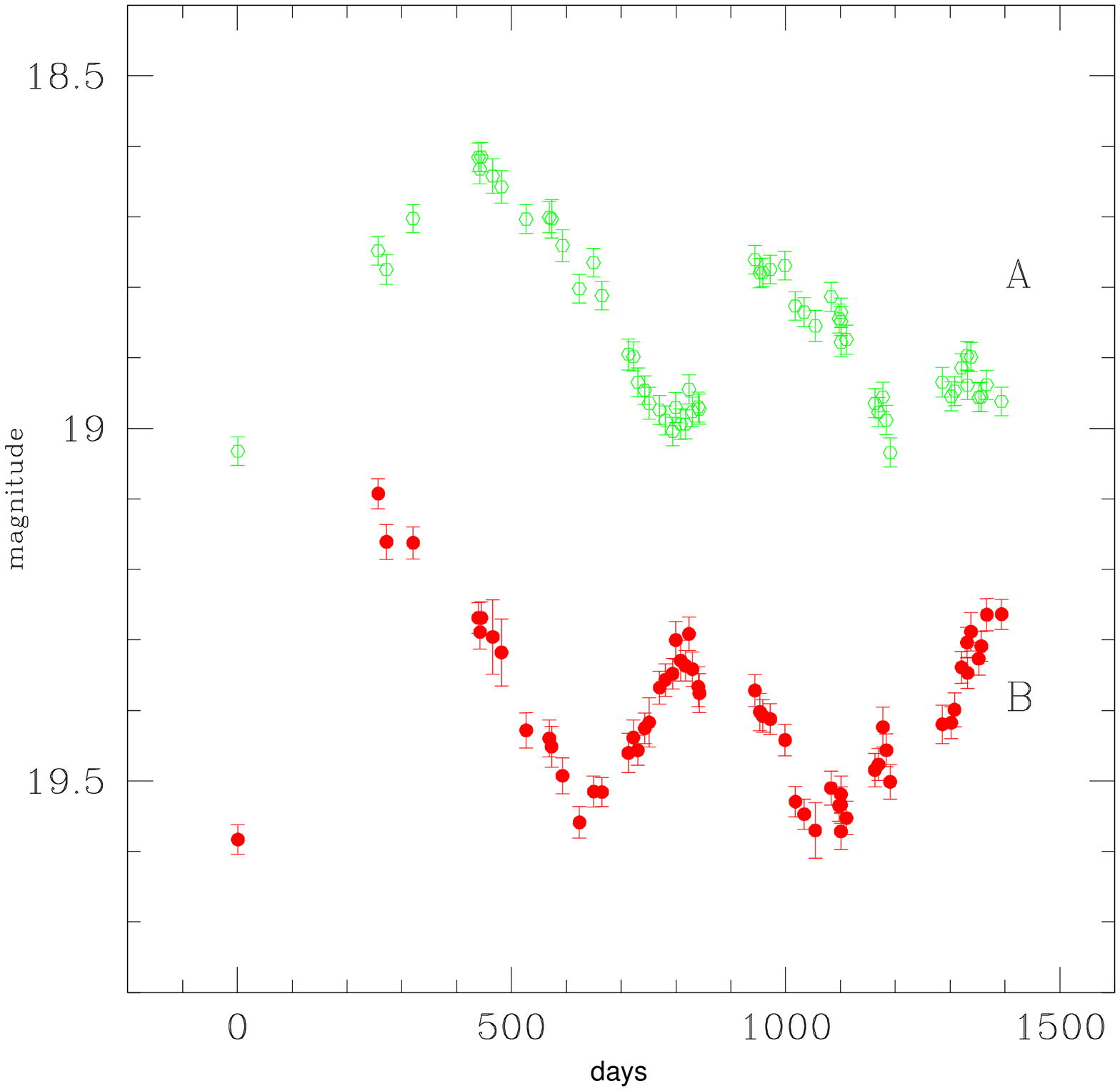}
\includegraphics[width=7cm]{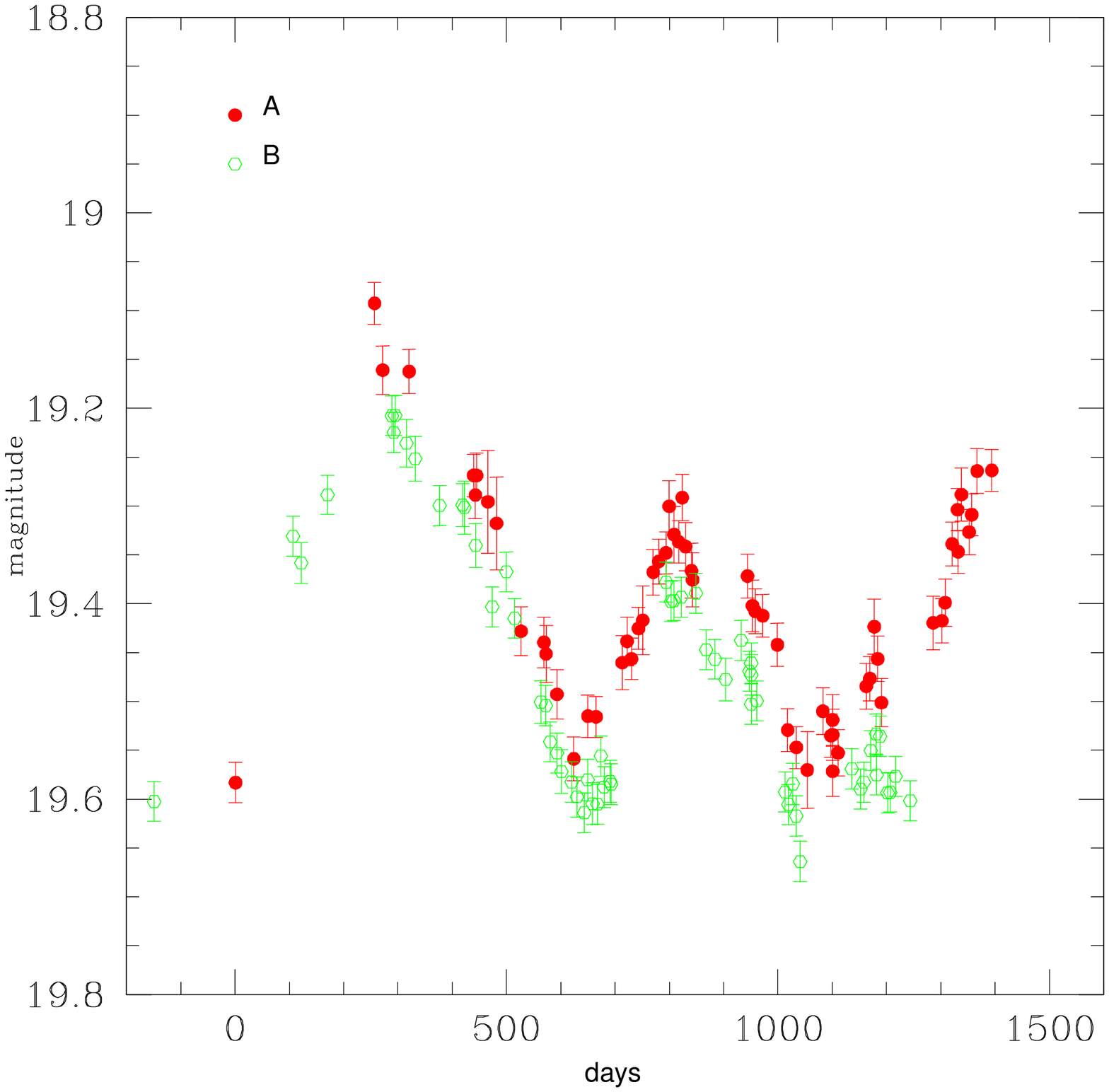}
\includegraphics[width=7cm]{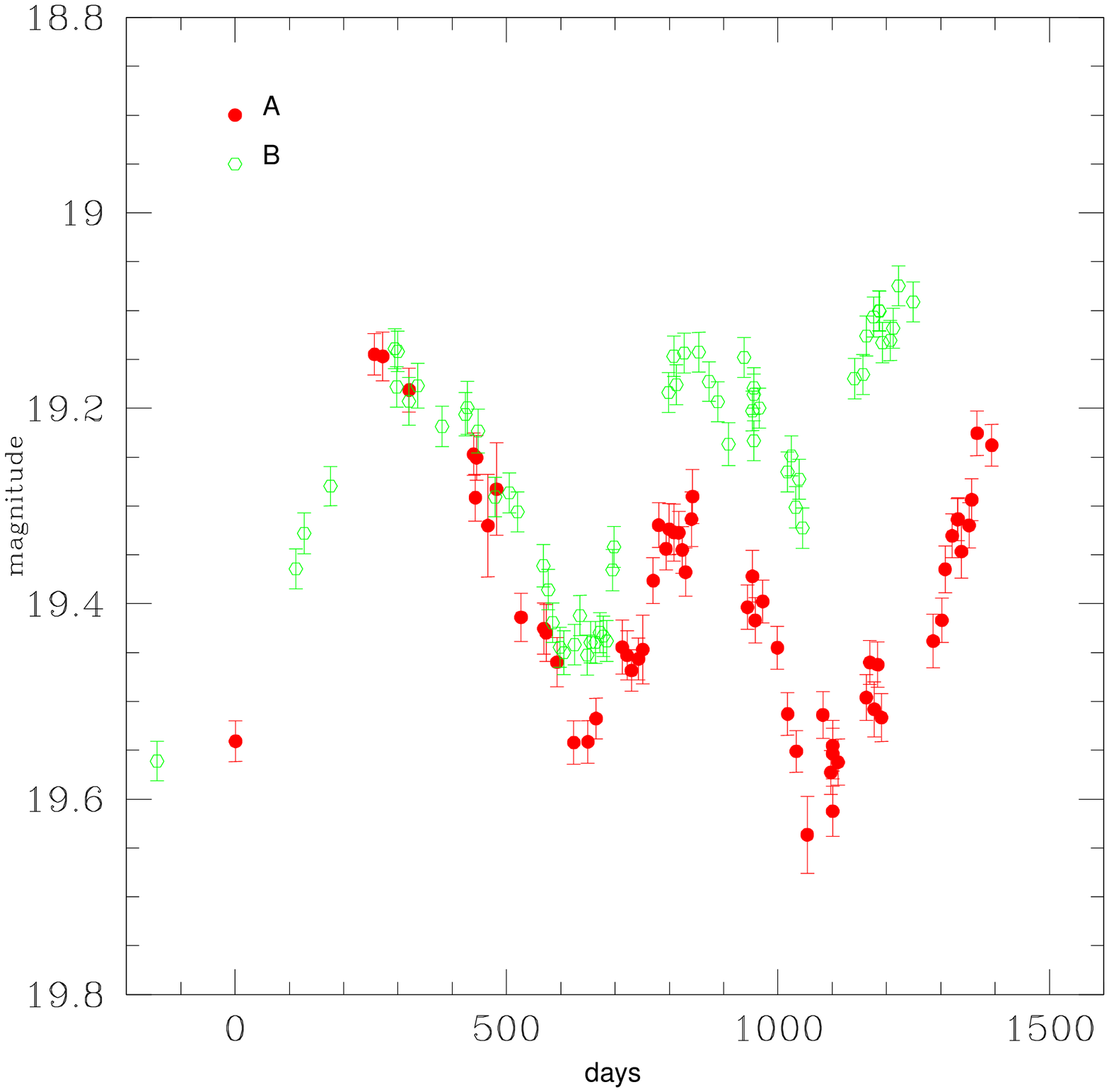}
\caption{{\it Top}: The simulated light curves of two quasar components 
in set 1a (see text and Table~\ref{results}). {\it Middle}: The time delay and magnitude shifted curves
in set 1b. The remaining difference between the two curves 
corresponds to the linear microlensing variation. 
{\it Bottom}: The light curves in set 1c, shifted in time delay
and magnitude. The remaining difference between the two curves is the
higher order  microlensing variation. }
\label{curv1}
\end{figure}

\begin{figure}
\includegraphics[width=7cm]{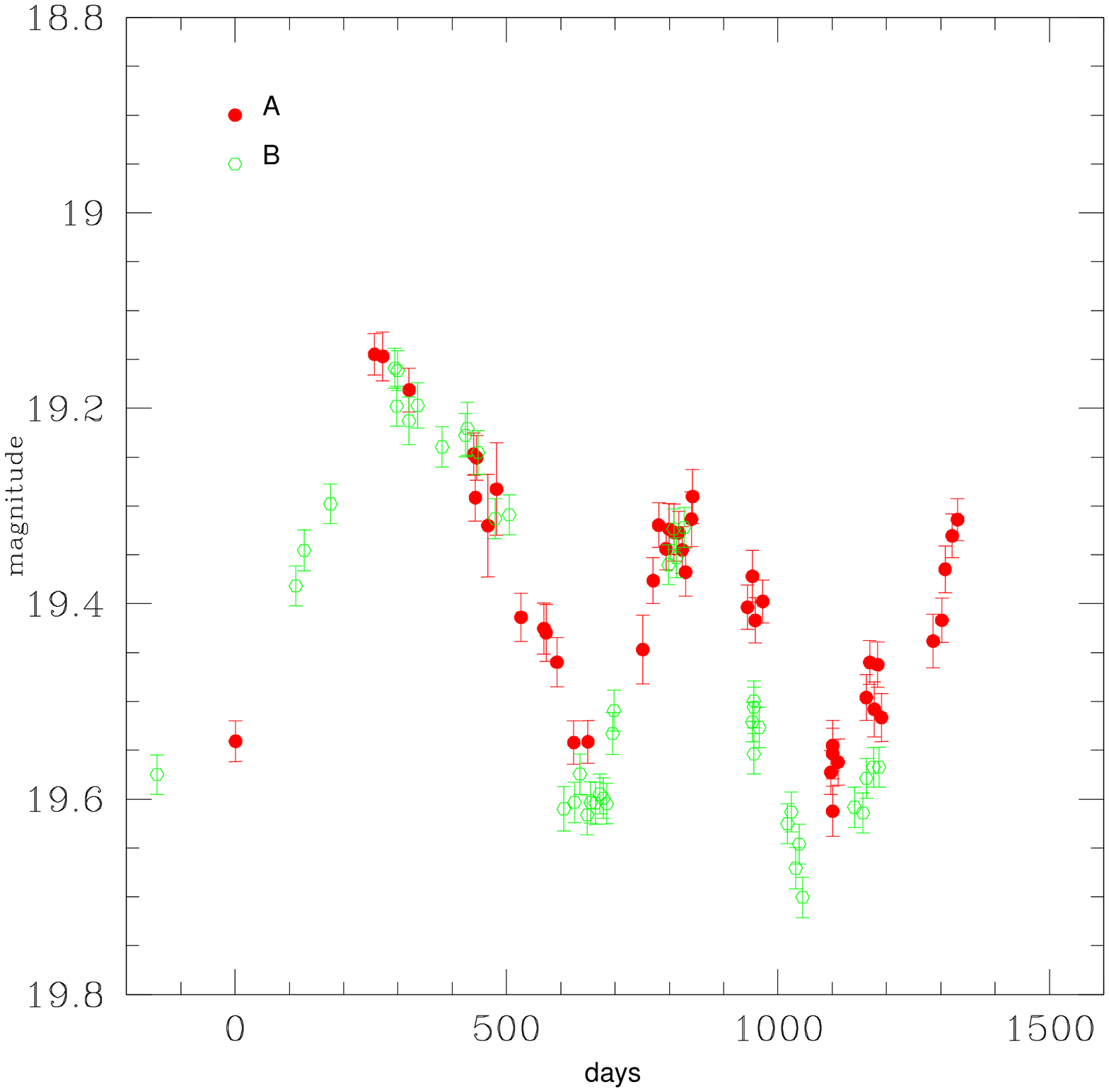}
\includegraphics[width=7cm]{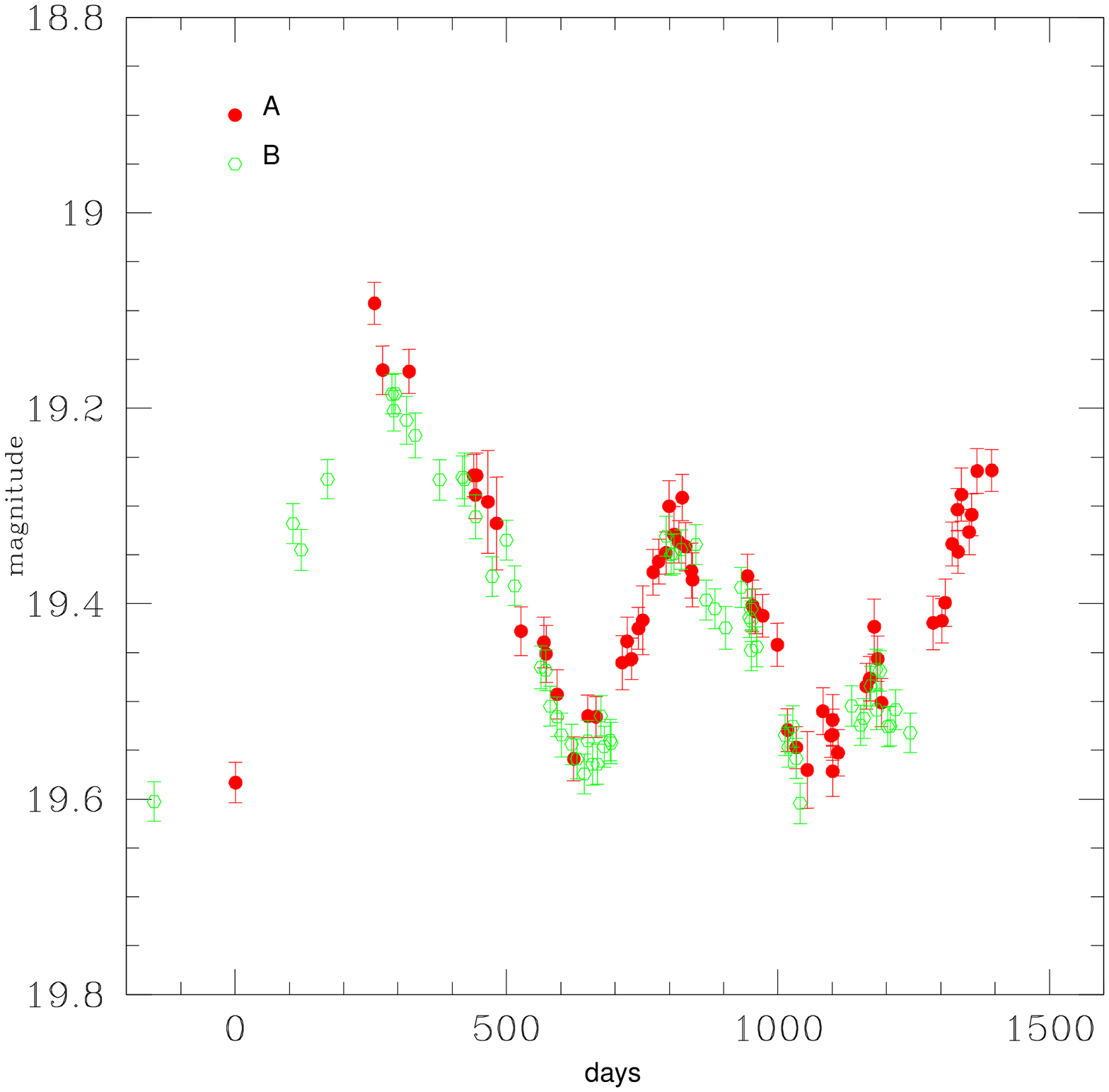}
\caption{{\it Top}: The time delay and magnitude shifted light curves form set 1a. 
{\it Bottom}: The microlensing-corrected light curves derived by the
iterative run on the curves in set 1c.
}
\label{curv1b}
\end{figure}

\begin{figure}
\includegraphics[width=7cm]{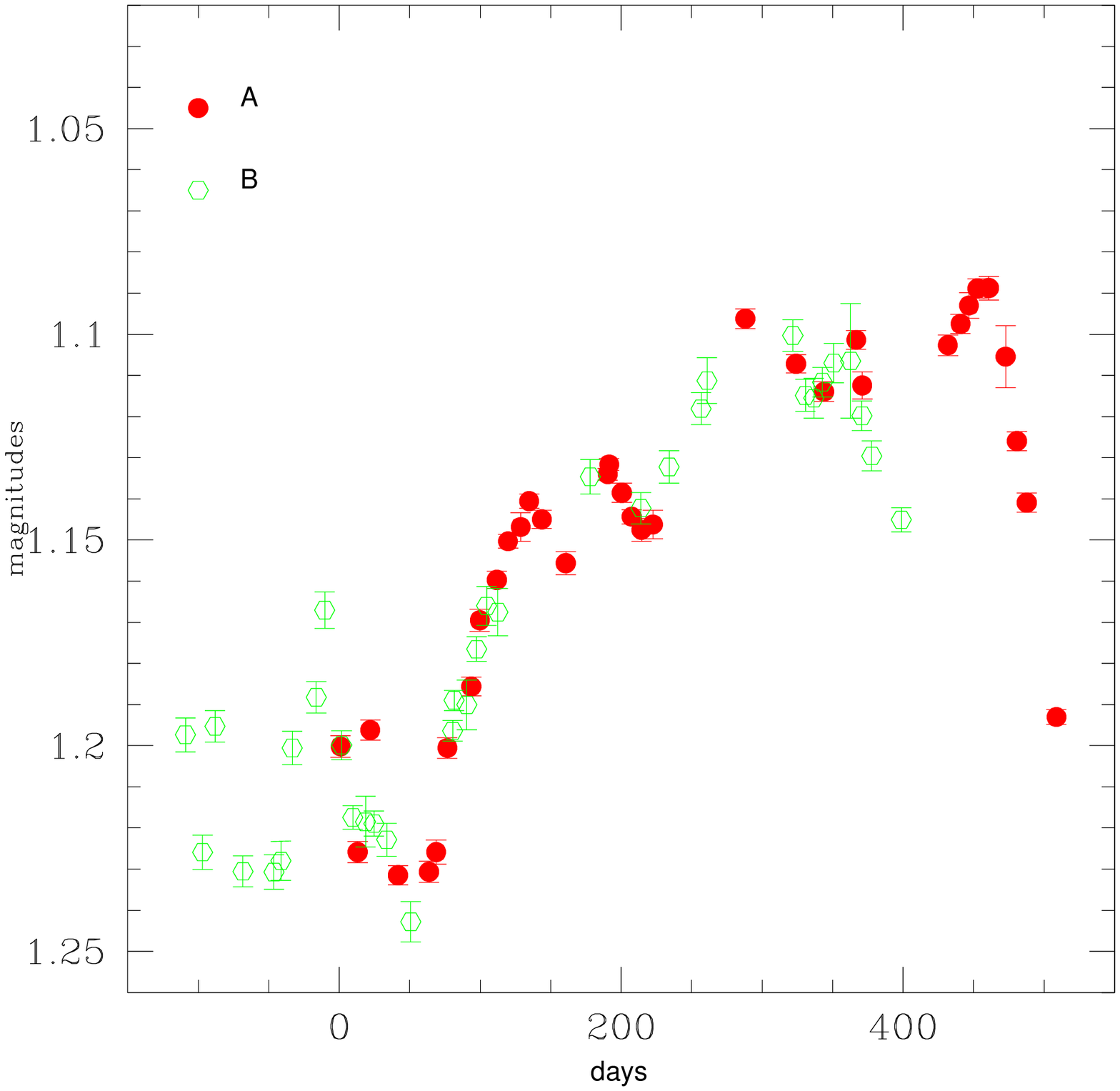}
\includegraphics[width=7cm]{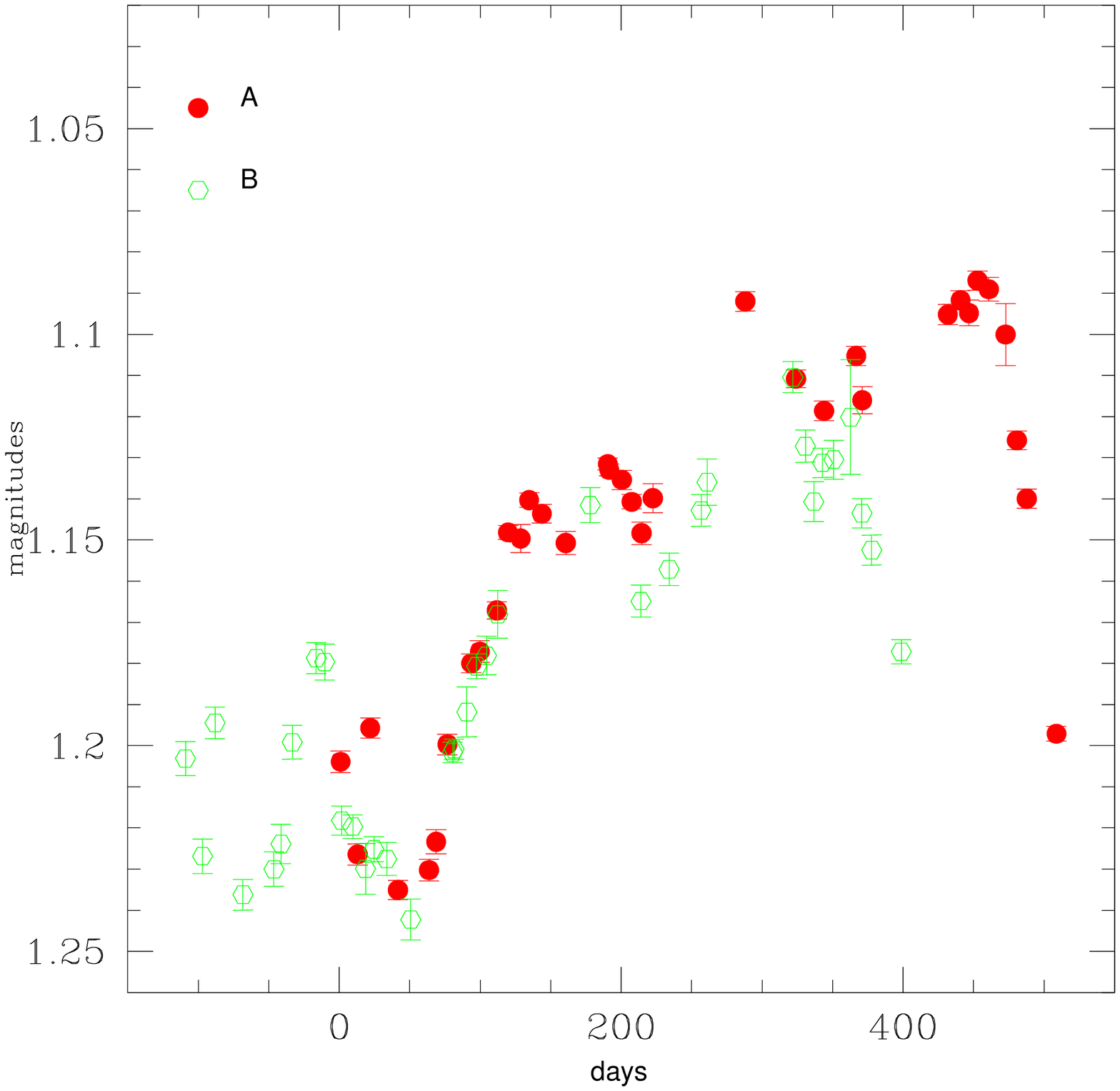}
\includegraphics[width=7cm]{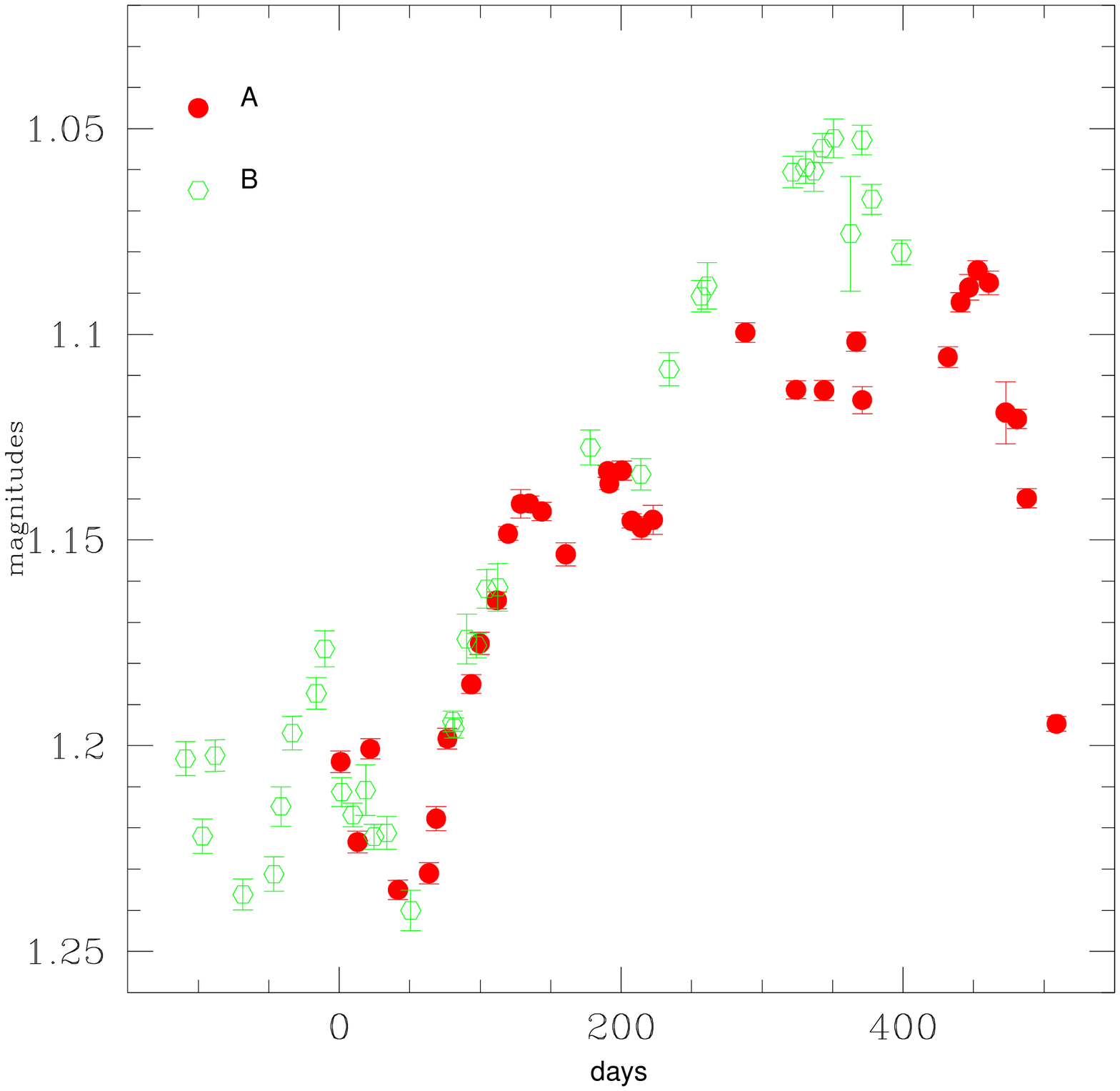}
\caption{{\it Top}:  The simulated light curves of two quasar components 
in set 2a (see text and Table~\ref{results}).
{\it Middle}: The time delay and magnitude shifted curves
in set 2b. The remaining difference between the two curves
corresponds to the
linear microlensing variation. 
{\it Bottom}: The light curves in set 2c, shifted in time delay
and magnitude. The remaining difference between the two curves is the
higher order  microlensing variation.}
\label{curv2}
\end{figure}

We apply the  method to simulated light curves  in order to understand
and estimate  the errors on  the measured values. Various  real quasar
light  curves were  used as  models  for typical  time variations  and
sampling of  data points. The measurement errors in the curves represent 
simulated photon noise errors, so that the faint component will have larger
errors than the bright one.  We present the  results from 
two sets of  simulated light  curves.  The first set
(1a, b  and c in Table~\ref{results})  consists of two curves,  A and B,
each of  60 points,  dispersed over  a time interval  of 2  years (see
Fig.~\ref{curv1} (top) and \ref{curv1b} (top)).  
The B curve is shifted by 145 days in time and
has a magnitude offset of 1.95 mag.  In set 1b we have added a slow
linear time  dependent variation with a slope $\alpha$ 
(Eq.~\ref{equation:min2})  in the  B curve in  order to simulate
long-time-scale  microlensing  effects  (Fig.~\ref{curv1}  middle).  A
higher order microlensing variation ($\alpha t + \beta t^2$) is 
added  to the B curve in set 1c
(Fig.~\ref{curv1} bottom).  In the second set (2a, b, and c) we have simulated
another kind  of intrinsic variation  and sampling.  The  curves span
an interval of two years  and they have 36 data points each.  The
time delay is 110 days and  the magnitude offset is 0.67 mag (see
Fig.~\ref{curv2} top).   In sets 2b and c, we added  linear (b) and
higher  order (c)  variations  in the  A  curve (see  Fig.~\ref{curv2}
middle and bottom).

\subsection{Time delay measurements}

The $\chi^2$ algorithm was applied  to all the simulated sets of light
curves.  The fit was performed several times in order to check for the
influence of various parameters such as: (i) the number of data points
in  the model curve  ($M$ in  Eq.~\ref{equation:min2}), (ii)  the FWHM
($\tau$) of  the Gaussian  used to smooth  the model curve  ($r(t)$ in
Eq.~\ref{equation:min2}), (iii) the FWHM ($2\sqrt{2ln(2)}\tau_{2}$) of
the  Gaussian that defines  the time  scale of  the variations  in the
curve (used to determine the weight $W_i$ in Eq.~\ref{equation:min2}),
and     (iv)     the      Lagrange     parameter     ($\lambda$     in
Eq.~\ref{equation:min2}).   For the  first set  the best  results were
obtained  when including a  weight ($W_i$  in Eq.~\ref{equation:min2})
whereas for the second set  the best results were obtained with no 
weight.  This  is because the data  points in the second  data set are
more regularly  distributed over time.  The  smoothing parameters were
chosen  so  that the  model  curve is  smooth  over  time, but  still
allowing variations  large enough  to fit the  data.  The  results are
shown in Table~\ref{results}.

We performed Monte Carlo simulations  to estimate the errors in the results.
Sets  of  1000  light  curves  were  simulated  with  error bars 
determined  randomly  from a  Gaussian  distribution  with a  $\sigma$
equal to the  measurement errors ($\sigma_{a}$ and
$\sigma_{b}$).   One  set was  made  with  the  same sampling  as  the
original curves (the  simulated examples) and another  set of 1000
curves  was created with  the same  number of  sampling points  as the
original, but randomly  distributed.

When applying the iterative algorithm it is natural to think that
the choice of the number of individual parts $n$ is important for the
results. Various tests on our simulated curves indicate that if  $n$ 
is too small, there is little  difference from the total curve
and higher order microlensing effects 
can not be  properly corrected for. If on the other hand $n$ is
too high, each part will contain very few data points and for some
parts (depending on the sampling) the overlap between the two curves 
will be too small to determine the external variations.  
The choice of $n$ will thus depend on the number of data points
and on how they are distributed  (e.g., presence of gaps in the curves). 
The optimal values for the curves in the
simulated curves, i.e., the values resulting in the best convergence
of $\Delta t $,  were found to be 3 and 2 in sets
1 and 2 respectively. 
The results shown in Table~\ref{results} correspond to the
mean and standard deviation of the values obtained.

\subsection{Results}

All the time delay measurements obtained in sets 1a, b and 2a, b are
very    well   determined   within    the   estimated    errors   (see
Table~\ref{results}).  As expected, the results depend on the sampling
of  the light  curves and  the error  estimates from  the  Monte Carlo
simulations  with  random  sampling  are  larger  than  the  ones
obtained with  a fixed sampling.   In sets 1b  and 2b where  a slow
linear variation is added in one  of the curves, the time delay is still well
determined whereas in sets 1c and 2c where higher order variations
are  added  there  are  small  systematic errors  in  the  time  delay
estimated with the direct method.  In these cases, 
the iterative method gives more  accurate results with
smaller uncertainties. Fig.~\ref{curv1b} displays how the microlensing
effects in data set 1c are corrected for with the iterative method.
In the sets with no microlensing (1a and 2a) we note that the
iterative method yields larger errors than the direct method. 
When no external variations (simulated microlensing effects) are present 
the best constrained time delay
value is obtained by setting $\alpha=0$, hence removing one of the
free parameters.  
The iterative method should therefore preferably only be applied where 
external variations are clearly present.

The magnitude shifts  between the two curves are  very well determined
in sets 2a, b  and c whereas in example 1a, b and  c there are small
systematic errors of $0.01-0.05$ mag.

\begin{figure}
\includegraphics[width=8cm]{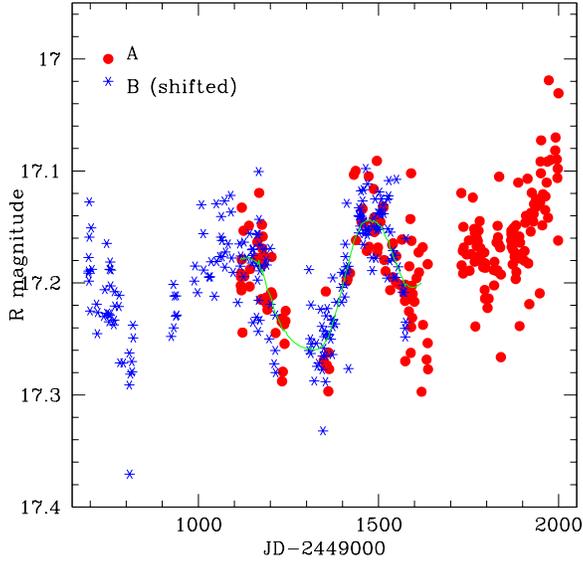}
\caption{The R-band light curves of QSO~0957+561.  The  curve is shifted 
by 423 days and 0.0625 mag. The solid line shows the model curve
derived by the $\chi^2$ algorithm.}
\label{b0957}
\end{figure}

\begin{table*}[t]
\begin{center}
\caption[]{{Results from the analysis of the simulated light curves. 
The sets (a, b and c) correspond to simulations with respectively 
no, linear and higher order external variations in one of the components.
These variations simulate microlensing effects (see the text for more details). }}
\begin{tabular}{lcccccc}
\hline
light curves                  &   1a &  1b & 1c & 2a & 2b &  2c \\
\hline 
true $\Delta t $  &      145     &     145    &  145       &  110       &   110      &   110      \\
\underline{derived $\Delta t$:} & & & & & & \\
direct method &      143$\pm$5     &     151$\pm$5    &  132$\pm5$        &  109$\pm$3       &   119$\pm$5      &   119$\pm$4\\ 
M.C. fixed sampling    &   146$\pm$5  & 148$\pm$5  & 135$\pm5$  &  109$\pm$3  &  114$\pm$5   & 115$\pm$4\\
M.C. random sampling   &   155$\pm$23 & 149$\pm16$           & 147$\pm$18 & 118$\pm$35 &   117$\pm$32  & 118$\pm$24\\
Iterative method  &  145$\pm$7         &  145$\pm2$&  143$\pm9$  &   108$\pm$6      &  111$\pm$3   & 114$\pm$8  \\ 

\hline
true $\Delta m $  &     1.950    &  1.950     &   1.950    &   0.670     &    0.670  &  0.670   \\ 
\underline{derived $\Delta m$:}  & & & & & & \\
direct method &     1.923$\pm$0.002    &  1.968$\pm$0.004     &   1.868$\pm0.004$    &   0.667$\pm$0.001    &     0.667$\pm$0.003  &  0.647$\pm0.004$   \\
M.C. fixed sampling    &1.927$\pm$0.002 & 1.964$\pm$0.004 & 1.885$\pm0.004$ & 0.667$\pm$0.001 & 0.665$\pm$0.003 &  0.678$\pm0.004$\\
M.C. random sampling   &1.921$\pm$0.005 & 1.963$\pm$0.036    & 1.905$\pm$0.08 & 0.672$\pm$0.017 & 0.65$\pm$0.04 & 0.65$\pm$0.02 \\ 
Iterative method  &  1.90$-$2.0      &  1.91$-$2.01  & 1.95$-$2.31  & 0.64$-$0.67 &   0.65$-$0.69          & 0.65$-$0.69\\ 
\hline
\end{tabular}

\label{results}
\end{center}
\end{table*}

\section{Application to light curves of QSO~0957+561}
\label{sect:b0957}

Our  method was  also  applied to a  public  data set  of
QSO~0957+561  published by Serra-Ricart  et al.  (\cite{Serra-Ricart})
(see Fig.~\ref{b0957}).   The R-band lightcurves contain 214 data points
and span the time interval from February  1996 to July  1998. We
removed  eight points  that   deviate by  more  than  0.05 mag  from
neighbouring  points  in both  components.   We  applied our  $\chi^2$
algorithm to the data and, as was done for the simulated data, we estimated
the error  bars using Monte  Carlo simulations. A time  delay estimate
$\Delta  t  =  423\pm9$  days  and  a  magnitude  shift  $\Delta  m  =
0.063\pm0.007$  were obtained.
Considering the two gaps in the data set, JD 2450242 to  2450347 
and JD 2450637 to  2450729, Serra-Ricart et al. have divided the data
set into two parts, DS I and DS II. Since the second part (DS II) show the 
largest activity
they have determined a time delay value on DS II only, corresponding
to the autumn 1997/spring 1998 season.  They measure a time delay  value of 
425$\pm$4 days  and a magnitude shift of  0.06 mag on these curves.
Measurements with two other methods, the dispersion spectra and discrete
cross-correlation techniques give values of $426\pm12$ and $428\pm8$ days
respectively on the same data set (Serra-Ricart et al.  1999). 
In order to compare the results we therefore also apply our method
to DS II.   On the curves in DS II we measure $\Delta  t  =  424\pm22$.
Both time delay values, the one measured on the entire data set and
the one measured on DS II, are compatible with the results from
Serra-Ricart et al. However our errors are larger than the ones
found by Serra-Ricart et al. 
In particular we note that our errors are considerably larger when
using only DS II than when using the entire data set.
Serra-Ricart et al. do not give any time delay measurement from the
total curve but their  method seems to 
be well suited for  time delay measurements on a continuous curve (DS II), 
even when the time span of observation is only of the order of the time delay
value.  With our method the
time delay is better constrained with longer time series,
even if gaps are present in the curves.

An additional reason for the differences in the results  might be that 
whereas we have discarded 8 points form the total light curve, 
Serra-Ricart et al. have discarded 23 points.
Given our limited knowledge of the data points we found no reason to exclude
more points.

\section{Conclusion}

We have developed a method to determine time delays between light rays
from  lensed quasar  images from  their respective  light  curves.  As
other $\chi^2$ based  methods (e.g., Press et al. \cite{Press}  and Barkanna
\cite{Barkanna}) we  assume that the  statistics of the measurement  errors are
Gaussian, and  we require  smoothness for nearby  points in  the light
curves. In our  method we have included an optional weight to each
of the data points, depending on the relative distance to neighbouring
points.   In this way  one point  isolated in  time will  receive more
weight than each individual point in a cluster of nearby points.

Since microlensing effects are  often encountered in real light curves
of lensed quasars  we have adapted our method in  order to correct for
such variations.   A linear term in  one of the curves  is included to
correct  for slow  long-term variations.   Realistic  simulations show
that the time  delay and the magnitude shift  between two light curves
are  determined to  within a  few percent  for cases  with no  or slow
microlensing effects.  For faster microlensing variations we find that
running the algorithm  in an iterative way yields  better time delays.
This  confirms what  was proposed  in  the time  delay measurement  of
B1600+434 (Burud  et al. \cite{Burud})  for which four methods  were 
applied: the minimum dispersion method 
(Pelt \cite{Pelt}), the SOLA method (Pijpers \cite{Pijpers94}, \cite{Pijpers}
), the method described
in the present paper and its iterative version.
The iterative method yielded the best constrained time delay in this
data set due to external variations in one of the components.

We also  applied our method to  a public data set  of QSO~0957+561 and
have shown that  our results are in agreement  with the published time
delay.

\begin{acknowledgements}
IB and SS are supported by  contract
 P\^ole d'Attraction  Interuniversitaire,
P4/05 \protect{(SSTC, Belgium)}. 
JH is supported  by  the Danish Natural  Science Research  Council  (SNF).
\end{acknowledgements}

\end{document}